\documentclass[prl,twocolumn,showpacs]{revtex4}

\usepackage{graphicx}
\usepackage{epsfig}
\usepackage{bm}
\usepackage{latexsym}

\begin{document}

\title{Statistics of Spectra for    
One-dimensional Quasi-Periodic Systems 
at the Metal-Insulator Transition}

\author{Yoshihiro Takada,$^{1}$ Kazusumi Ino,$^{1}$ 
and Masanori Yamanaka$^{2}$}

\affiliation{
$^1$Department of Pure and Applied Sciences, University of Tokyo, 
Komaba 3-8-1, Meguro-ku,Tokyo, 153-8902, Japan \\
$^2$Department of Physics,
College of Science and Technology, Nihon University, \\
Kanda-Surugadai 1-8,
Chiyoda-ku, Tokyo, 101-8308, Japan}

\begin{abstract}
We study spectral statistics  of one-dimensional 
quasi-periodic systems at the metal-insulator  transition.  
Several types of spectral statistics are observed 
at the critical points, lines, and region.  
On the critical lines, we find the bandwidth distribution $P_B(w)$ 
around the origin (in the tail)
to have the form of 
$P_B(w) \sim w^{ \alpha}$ ($P_B(w) \sim e^{ -\beta w^{\gamma}}$) 
($\alpha , \beta , \gamma > 0 $),
while in the critical region $P_B(w) \sim w^{-\alpha'}$
($\alpha' > 0$).
We also find the level spacing distribution to follow 
an inverse power law $P_G(s) \sim s^{- \delta}$ ($\delta > 0$ ).

\end{abstract}

\pacs{05.45.Mt, 03.65.Sq, 71.30.+h, 71.23.An}

\maketitle

The universality of statistical properties of energy spectrum  
is ubiquitously found in quantum physics \cite{guhr}, 
ranging from quantum chaos \cite{bohigas} 
to quantum chromodynamics 
\cite{verbaarschot}.
In condensed matter physics, 
the metal-insulator (MI) transition 
in disordered electron systems gives a notable example 
of such universality.
In the metallic side, 
the energy level statistics can be described 
by random matrix theory \cite{dyson}.
The level spacing distribution is close to the Wigner surmise.
The universality class is classified by the symmetry of the system.
In the insulating side the level spacing distribution is Poissonian.
At the critical point, new statistical features
emerge \cite{altshuler,shklovskii}.
They are different from random matrix theory and Poisson statistics.
The universality of these new  statistics dubbed
as ``critical level statistics'' is also classified
by symmetry of the ensemble
\cite{aronov,hofstetter, kawarabayashi,cuevas}.

Quasi-periodic systems are other interesting ones
showing the MI transition.
The universality of the transition  has been characterized
by multifractal structures of bandwidths and wavefunctions
(see Ref. \cite{hiramoto} for a review).
The Harper model \cite{azbel,aubry}
is obtained by a gauge fixing of two-dimensional electrons
on the square lattice in a uniform magnetic field.
Especially incommensurate limits of the flux per plaquette,
such as the inverse of the golden mean $\sigma=\frac{-1+\sqrt{5}}{2}$,
have been extensively studied.
In the atomic scale, the penetration of flux for each plaquette
requires an enormous strength of magnetic field,
but nowadays it is realized in laboratory
on a quantum dot lattice
\cite{vonkli}.

The quasi-periodic system can be seen
as a system between ``periodic'' and ``random,''
thus giving an interesting example of quantum ``chaos.''
Classically an open orbit (separatrix) 
of equi-energetic curve appears
at the MI transition point.
But the quantum transition cannot be understood
by the classical consideration.
It is not characterized by a single energy band corresponding
to the separatrix but 
the wavefunctions for all the energy bands
become critical with power-law decay.
The energy level statistics are different from disordered systems.
It is not close to the Wigner surmise on the metallic side
nor is the Poisson statistics seen on the insulating side
\cite{megann}.
Spectra of quasi-periodic systems
have a fractal structure \cite{azbel},
which hinders established techniques used in disordered case
such as ``unfolding" to extract universal properties of fluctuations.

Recently Evangelou and Pichard \cite{evangelou} investigated
the bandwidth distribution 
$P_B(w)$ ($w$ : bandwidth)
of the one-dimensional
Harper model at the incommensurate limit of $\sigma$.
They found that
it agrees with
the semi-Poisson statistics
$P_B(w)=4w e^{-2w}$ 
with the sub-Poisson number variance
$\Sigma_2(E) \sim const. +\chi E$ ($\chi <1$)
at the critical point.
They argued that it is related to the universal characterization
of quantum chaos.

Although this suggests that a characterization 
by level-statistical idea is possible for the MI 
transitions of a larger class of quasi-periodic systems,
the universal appearance of the semi-Poisson statistics may
be questioned from the experience in critical level statistics
of disordered systems.
In Ref.\cite{aronov},
it has been suggested the level spacing distribution may be
described by a more generalized form 
$P(s) = A s^{\alpha}e^{-\beta s^{\gamma}}$
$(\alpha,\beta,\gamma > 0)$,
and it has been observed in numerical studies 
\cite{hofstetter}.
A similar rich structure of critical level statistics should be
expected for quasi-periodic systems.

Motivated by this idea,
we investigate a generalization of the Harper model,
namely the extended Harper model, which is obtained
from two dimensional electrons on the square lattice 
with next-nearest-neighbor hopping in a uniform magnetic field
\cite{thouless,han-thouless}.
We find that such a rich structure indeed emerges 
for the critical level statistics of this extended model,
but with some substantial modifications,
even for the case treated in \cite{evangelou}.
We explain this variety of critical level statistics 
by the quantum nature of the system.

\begin{figure}
  \begin{center}
    \epsfxsize=8cm
    \epsfbox{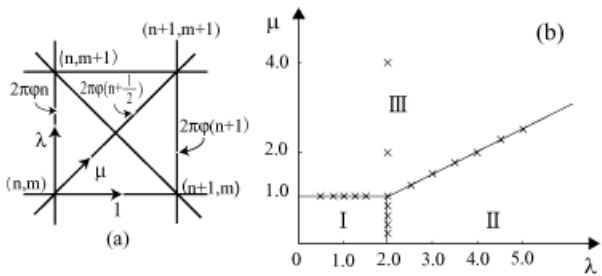}
    \caption{
 (a) Transfer integrals of the extended Harper model (Eq.(\ref{hamiltonian}))
      and (b)  
      its phase diagram 
      (See \cite{han-thouless}).
      In region I 
      the wavefunction (spectrum) is extended 
      (absolutely continuous)
      and in region I\hspace{-.1em}I
      it is localized (pure points).
      In region I\hspace{-.1em}I\hspace{-.1em}I and
      on the three boundary lines, 
      it is critical (singular continuous).
    }
    \label{fig;phase_lattice}
  \end{center}
\end{figure}

Let us first describe the model.
We consider the two-dimensional fermions 
on the square lattice 
in a flux per plaquette $\varphi$ 
with next-nearest-neighbor coupling 
(Fig.\ref{fig;phase_lattice}(a)). 
The Hamiltonian is given by 
\begin{eqnarray}
H=\cos p_x + \lambda \cos x + \mu \cos (p_x+x)
+\mu \cos (p_x-x),
\label{hamiltonian}
\end{eqnarray}   
where $p_x=-2\pi i \phi \frac{d}{dx}$.
The positive parameters, $\lambda$ and $\mu$, 
are the strength of the kinetic terms for the perpendicular
and next-nearest-neighbor directions respectively,
while the horizontal one is normalized to unity.
We take the Landau gauge $\Psi(x_n,y_n) = \psi_x(x_n) e^{i \nu y_n}$
and obtain the Schr\"odinger equation for $\psi_n = \psi_x(x_n)$ as
\begin{eqnarray}
\left[ 
1+ \mu \cos \left(2\pi \big(n + \frac{1}{2}\big)\varphi+\nu \right) 
      \right] \psi_{n+1}    \nonumber \\
+\lambda \cos (2 \pi  n \varphi  + \nu)  \psi_{n}   \nonumber \\   
+\left[
1+ \mu \cos \left(2\pi \big(n - \frac{1}{2}\big)\varphi + \nu \right)
      \right] 
\psi_{n-1}   
    &=& 
E \psi_{n}.
\label{eq:ex_harper_model}
\end{eqnarray}
At $\mu=0$,
this is reduced to the Harper model. 
For a rational $\varphi=p/q$ 
where $p$ and $q$ are relatively prime,
 Eq.(\ref{eq:ex_harper_model}) reduces 
to a $ q \times q $ matrix
with the boundary condition $\psi_{n+q}=\exp(ikq)\psi_{n}$,
which can be transformed to be tridiagonal \cite{thouless}.
To investigate the irrational limit of $\varphi \to \sigma$, 
we use the rational approximation of $\varphi_j=F_{j-1}/F_j$ 
where $F_j$ is the Fibonacci number satisfying 
$F_{j+1}=F_j+F_{j-1}$ with $F_0=1$ and $F_1=1$.

The phase diagram is shown in 
Fig.\hspace{-.2cm} \ref{fig;phase_lattice}(b).
The scaling property of the bandwidth 
was studied in \cite{han-thouless}. 
For large $q$, the minimum and maximum are obtained
at either $k=0$ or $\pi/q$ for fixed $\nu$ \cite{thouless}. 
We numerically diagonalized the matrices there
setting $\nu=0$ for simplicity,
and obtained the bandwidth distribution $P_B(w)$.
The normalizations are 
$\int^{\infty}_0 P_B(w) dw =1$ 
and $\langle w \rangle = \int^{\infty}_0 w P_B(w)dw=1 $.

\begin{figure}
  \epsfxsize=8cm
  \epsfbox{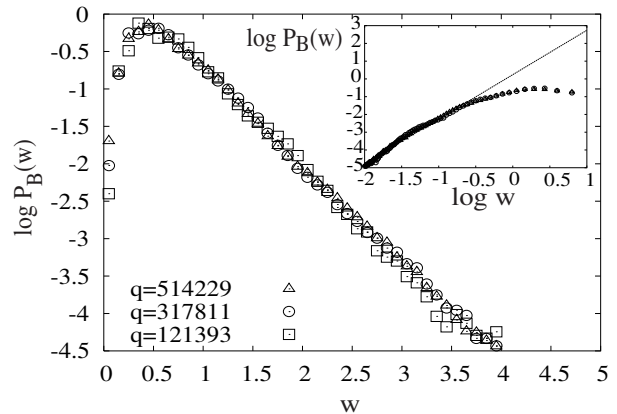}
   \caption{
     The bandwidth distribution 
     at $( \lambda , \mu ) = ( 2.0 , 0.4 )$,
     and 
     the one near the origin (inset).
   }
   \label{fig;large_l2}
\end{figure}

First, we study the bandwidth distribution
along the critical line $\lambda = 2$.
In Fig.\ref{fig;large_l2}, 
we plot the $P_B(w)$ at $( \lambda , \mu )=( 2.0 , 0.4 )$
for $j=25, 27, 28$.   
It shows a good convergence to a limit distribution,
indicating the existence of the limit of the 
bandwidth distribution at the incommensurate flux $\varphi$.
For $0 \leq \mu < 1$ ,
we find $P_B(w)$ exhibits a continuous change.
It has been concluded \cite{han-thouless} that the 
scaling property 
of the bandwidths is invariant on this line, 
implying the system belongs to the same universality class.
The continuous change means
that $P_B(w)$ is not a universal quantity. 

\begin{figure}
  \epsfxsize=5.5cm
  \epsfbox{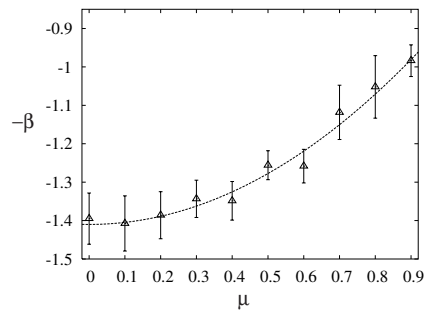}
  \caption{
    $\mu$-dependence of index $\beta$
    along the critical line $\lambda = 2$.
   }
   \label{fig;parabola}
\end{figure}

In Fig.\ref{fig;large_l2}, one sees linear behaviors 
at large $w$, implying an asymptotic form 
\begin{equation}
  P_B(w) \sim e^{-\beta w},\quad {\rm as}\quad  w \rightarrow \infty
  \label{tail}
\end{equation} 
where $\beta >0$.  
The optimized values of  $\beta$  are shown
in Table \ref{table;l2}. 
It suggests a quadratic dependence of $\beta$ on $\mu$.  
The best fit is attained by a quadratic function 
$ \sim  1.4 - 0.53 \mu^2$
(Fig.\ref{fig;parabola}).
This rather simple dependence may be related 
to the universality class on this line.   

$P_B(w)$ is estimated to be zero at the origin. 
Inset of Fig.\ref{fig;large_l2} 
shows $P_B(w)$ near the origin,
indicating a power law.
To characterize this behavior, we make an ansatz of
\begin{equation}
  P_B(w) \sim w^{\alpha}, \quad {\rm as } \quad w \rightarrow 0 
  \label{origin}
\end{equation}
where $\alpha >0$.   
The optimized values of  $\alpha(\mu)$ are shown 
in Table \ref{table;l2}.    
One sees that $\alpha(\mu)$s are 
stable, 
suggesting that $\alpha(\mu)$ 
is 
independent of $\mu$
at the incommensurate limit. 

These numerical results suggest that 
the overall distribution is given 
by a generalized semi-Poisson form 
\begin{equation} 
  P_B(w) = Aw^{\alpha} e^{-\beta w}
  \label{semi-Poisson}
\end{equation}
where $A$ is a normalization constant. 
The exact semi-Poisson form i.e. $\alpha = 1$, $\beta = 2$
has been reported at $(\lambda,\mu)=(2.0 , 0.0)$ \cite{evangelou}.
Actually the normalization conditions 
allow only one parameter $\alpha$ with $\beta=\alpha+1$ and 
$A=\frac{(\alpha+1)^{\alpha+1}}{\Gamma(\alpha+1)}$. 
It is clear that such relations are not consistent with 
our results as shown in Table \ref{table;l2}.
Especially the semi-Poisson statistics 
with $\alpha=1$ do not reproduce the index
of the power-law behavior around the origin 
at $\mu=0$.
Using one-parameter fit by (\ref{semi-Poisson}),
we get $\alpha \sim 0.7$ to the overall distribution,
which is much smaller than that estimated by (\ref{origin}).
For $\mu \neq 0$, 
we get similar deviations. 
Thus we may conclude that the semi-Poisson form for $P_B(w)$ 
is only an approximation to the overall distribution.

\begin{figure}
  \epsfxsize = 8cm
  \epsfbox{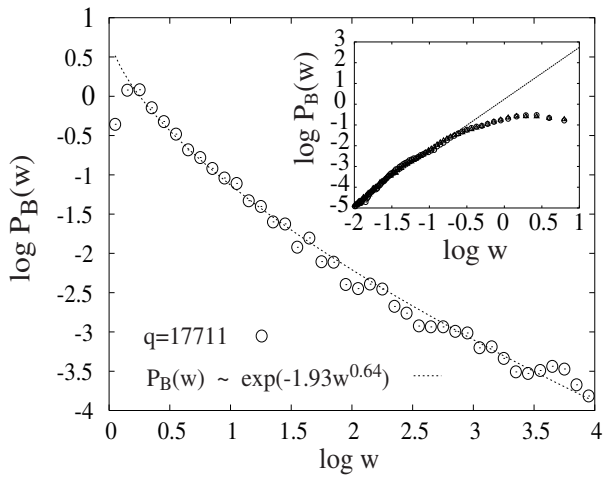}
  \caption{
   The bandwidth distribution at 
   $( \lambda , \mu ) = ( 1.0 , 1.0 )$,
    and the one near the origin (inset).
  }
  \label{fig;tailfit_m1}
\end{figure}

Next, we investigate $P_B(w)$ on the other critical lines. 
It turns out that 
the behaviors are different
from those on the critical line $\lambda = 2$.   
Fig.\ref{fig;tailfit_m1} shows the $P_B(w)$ 
at $( \lambda , \mu ) = ( 1.0 , 1.0 )$.
%
%
The large $w$ behaviors do not show an exponential decay,
but a milder one.
Thus we make a generalized ansatz of
\begin{eqnarray} 
  P_B(w) \sim e^{-\beta \omega^{\gamma}} 
  \quad {\rm as}\quad  w \rightarrow \infty
  \label{tail2}
\end{eqnarray}
with $0< \gamma <1$. 
The optimized curve agrees well with the obtained data
(Fig.\ref{fig;tailfit_m1}).
The behavior near the origin is also found 
to follow a power law (inset of Fig.\ref{fig;tailfit_m1}).
On the line $\mu = 1$,
the index $\alpha$ seems to be 
constant
around 1.5.
We find that 
the numerical data on the critical line $\lambda=2\mu$
can be also described 
by each ansatz (\ref{origin}) and (\ref{tail2}).
The index $\alpha$ is stable
around 1.5 as in Table \ref{table;not_l2},
which is near the values on the critical line $\mu = 1$.

The overall distribution on the critical lines 
$\mu=1$ and $\lambda = 2 \mu$ except for the bicritical point 
may be cast in a form 
\begin{eqnarray} 
  P_B(w) =A w^{\alpha} e^{-\beta \omega^{\gamma}}. 
  \label{ansatz}
\end{eqnarray}
This form of distribution was suggested 
in Ref.\cite{aronov} for 
the level spacing distribution of disordered systems 
at the MI transition.   
The $\gamma \neq 1$ case has been encountered 
in a 3D disordered system with orthogonal symmetry 
\cite{hofstetter}. 
The normalization conditions constrain 
$\beta=\left( 
\frac{\Gamma(\frac{\alpha+2}{\gamma})}
{\Gamma(\frac{\alpha+1}{\gamma})} 
)\right)^{\gamma}$ and $A=
\frac{\gamma \beta^{\frac{\alpha+1}{\gamma}}}
{\Gamma(\frac{\alpha+1}{\gamma})}$.  
Our analysis does not support these constraints.
Again, we tried a one-parameter fit varying 
$\alpha$($\gamma$)
in (\ref{ansatz}) with the observed value of 
$\gamma$($\alpha$) fixed.
We got values much smaller (larger) than
the observed indices $\alpha$ ($\gamma$) around the origin
(in the tail)
for all the points we studied on the critical lines $\mu=1$ and
$\lambda=2\mu$.

We also investigate the bicritical point at 
$( \lambda , \mu) = ( 2.0 , 1.0 )$. 
However, the convergence of the obtained distribution is slow, 
and we have not got information of the incommensurate limit. 

\begin{figure}
  \epsfxsize=8cm
  \epsfbox{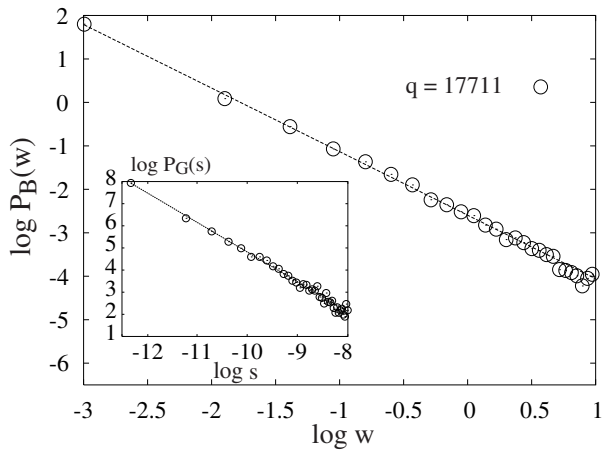}
   \caption{ The band width distribution and 
    level spacing distribution (inset)
    at $(\lambda , \mu) = (2.0 , 2.0)$.
   }
   \label{fig;gap}
\end{figure}

The values of $\alpha$ and $\gamma$ 
on the critical lines $\mu = 1$ and $\lambda = 2 \mu$
are considerably smaller from those on 
$\lambda = 2$ ($\gamma$=1 there). 
This smallness of  $\alpha$ and $\gamma$ 
indicates the tendency of $P_B(w)$ on these critical lines 
to  broaden both  
to the origin and the tail compared to $P_B(w)$ on $\lambda=2$.   
This may imply that the transport property 
becomes different when $\mu$ gets sufficiently large.
To supplement this observation,  
we investigate the bandwidth distribution 
in the critical region (the region III in Fig.\ref{fig;phase_lattice}(b)).  
$P_B(w)$ follows an inverse power law $P_B(w) \sim w^{-\alpha'}$ 
($\alpha' > 0$) there (Fig.\ref{fig;gap}).
The values of $\alpha'$ are summarized in Table \ref{table;l2}. 
This behavior sharply contrasts 
with the one seen on the critical lines.
The divergence of $P_B(w)$ at the origin implies 
the dominance of relatively flat bands, 
indicating that in most of the bands
the wavefunction is   close to be localized.
Also the power law decay in the tail means 
the appearance of bands whose wavefunction is
close to be extended.

From the two-dimensional point of view, 
the next-nearest-neighbor hopping term 
($\mu$ dependent term) is dominant
in Eq.(\ref{hamiltonian}) 
in the critical region. 
Since $y$-translation is canonically conjugate 
to $x$-translation in the uniform magnetic field,   
the extension (localization) of the wavefunction in $y$-direction must be 
balanced by the localization (extension) in $x$-direction to satisfy 
the uncertainty principle. 
When $\mu $ is small, $x$-dependence of 
the wavefunction is determined  by $\lambda$, 
the anisotropic parameter for $y$-direction.  
On the other hand, when $\mu$ is sufficiently large, 
as $\mu$ acts on $x$- and $y$-directions equally,
the wavefunction is  sensitive to $\mu$ by both of 
its $x$- and  $y$-dependences. This makes the 
MI transition in the critical region different 
from that on the critical line $\lambda=2$,  
resulting in 
the increase of the bands close to be localized 
as well as the bands close to be extended. 
Thus the appearance of the inverse power law in the critical region 
and the differences of $\alpha$ and $\gamma$ 
among the critical lines
are consequences of quantum nature of the system. 

We also investigate the gap distribution $P_G(s)$. 
The gap distribution at $( \lambda , \mu) = ( 2.0 , 0.0 )$
has been known to follow 
an inverse power law \cite{machida,geisel},
which diverges at the origin 
\begin{equation}
P_G(s) \sim s^{-\delta}
\label{eq;gap_fit}
\end{equation}
with $\delta \sim 1.5$. 
We find $P_{G}(s)$ along the three critical lines 
and in the critical region also follows an inverse power law
(inset of Fig.\ref{fig;gap}). 
The values of $\delta$ are
shown in Table \ref{table;l2} and \ref{table;not_l2}.  

It would be interesting to examine a similar statistical 
distribution for the wavefunctions. 
The characterization through level-statistical idea 
may have a good possibility.  

In this letter,
we have seen that a systematic characterization by a 
level-statistical idea for the quasi-periodic system is possible.
We have investigated 
the bandwidth $P_B(w)$ and 
gap distributions $P_G(s)$ for one-dimensional 
quasi-periodic Schr\"odinger equations 
at the MI transition and confirmed their variety
as expected.
We found a power law $P_B(w) \sim w^{\alpha}$ 
near the origin and 
a generalized exponential decay $P_B(w) \sim e^{-\beta w^{\gamma}}$
($\alpha, \beta, \gamma > 0$)
on the critical lines,  
while in the critical region $P_B(w)$ follows an inverse power law 
$\sim w^{-\alpha'}$($\alpha' > 0$). 
We gave the explanation for this variety 
of $P_B(w)$ by quantum nature of the system.
A surmise of a form $P_B(w)=A w^{\alpha}e^{-\beta w^{\gamma}}$ 
gives only an approximation.
The gap distribution 
$P_G(s)$ shows an inverse power law  
$ \sim s^{-\delta}$ for whole the phase diagram.  
For the bicritical point, we did not get a conclusive result. 

\setlength{\arrayrulewidth}{0.8pt}
\begin{table}[tb]
\begin{tabular}{ccccc}     \hline
$\lambda$ & $\mu$ & $\alpha$ 
& $\beta$ & $\delta$  \\  \hline
%
%
%
%
%
%
2.0 & 0.0 & 

2.5($\pm$ 0.1) &
1.39($\pm$ 0.07) &
1.5($\pm$ 0.2)    \\

2.0 & 0.1 & 
2.5($\pm$ 0.1)&
1.41($\pm$ 0.07)  &
1.5($\pm$ 0.2)    \\

2.0 & 0.2 & 
2.5($\pm$ 0.1 ) &
1.39($\pm$ 0.06)  &
1.5($\pm$ 0.2)  \\

2.0 & 0.3 & 
2.51($\pm$ 0.08) &
1.34($\pm$ 0.05) &
1.5($\pm$ 0.1)   \\

2.0 & 0.4 & 
2.51($\pm$ 0.08) &
1.35($\pm$ 0.05) &
1.5($\pm$ 0.2)   \\

2.0 & 0.5 & 
2.50($\pm$ 0.06) &
1.26($\pm$ 0.04) &
1.5($\pm$ 0.2)  \\

2.0 & 0.6 & 
2.52($\pm$ 0.07) &
1.26($\pm$ 0.04) &
1.5($\pm$ 0.2)  \\

2.0 & 0.7 & 
2.54($\pm$ 0.07) &
1.12($\pm$ 0.07) &
1.5($\pm$ 0.2)  \\

2.0 & 0.8 & 
2.52($\pm$ 0.07) &
1.05($\pm$ 0.08)  & 
1.5($\pm$ 0.1)   \\

2.0 & 0.9 & 
2.48($\pm$ 0.04)&
0.98($\pm$ 0.11) &
1.47($\pm$ 0.04)   \\

2.0 & 1.0 & 
 - &
 - &
1.2($\pm$ 0.3) \\
%
%
%
%
%
%

2.0 & 2.0 &
-1.46($\pm$ 0.06)&
- &
1.34($\pm$ 0.07)     \\

2.0 & 4.0 &
-1.5($\pm$ 0.1)&
-  &
1.1($\pm$ 0.1)   \\

2.0 & 6.0 &
-1.37($\pm$ 0.08)&
- &
1.3($\pm$ 0.2)     \\
\hline
\end{tabular}
\caption{
The optimized indices on the critical line $\lambda = 2$ 
and in the critical region. 
For the definitions of $\alpha$, $\beta$, and $\delta$,
see 
Eqs.(\ref{tail}), 
(\ref{origin}), and
(\ref{eq;gap_fit}) respectively.
Also the value of $\alpha$ in the critical region 
 is defined as  $-\alpha'$.}
\label{table;l2}
\end{table}

%
%
%
%
%
%
%
%
\setlength{\arrayrulewidth}{0.8pt}
\begin{table}
\begin{tabular}{ccccc}     \hline
$\lambda$ & $\mu$ & $\alpha$
& $\gamma$ & $\delta$  \\  \hline

0.5 & 1.0 & 
1.55($\pm$ 0.14) &
0.41($\pm$ 0.04)&
1.29($\pm$ 0.08)    \\

0.75 & 1.0 & 
1.48($\pm$ 0.17) &
0.52($\pm$ 0.08)&
1.3($\pm$ 0.2)  \\

1.0 & 1.0 & 
1.48($\pm$ 0.12)&
0.65($\pm$ 0.07)& 
1.3($\pm$ 0.2)  \\

1.25 & 1.0 & 
1.61($\pm$ 0.17)&
0.53($\pm$ 0.05)&
1.4($\pm$ 0.2)  \\

1.5 & 1.0 & 
1.48($\pm$ 0.13) & 
0.60($\pm$ 0.07)& 
1.29($\pm$ 0.10)  \\
%
%
%
%
%
%
%
%
%

2.5 & 1.25 & 
1.44($\pm$ 0.10)&
0.54($\pm$ 0.06) &
1.31($\pm$ 0.10) \\

3.0 & 1.5 & 
1.61($\pm$ 0.13)& 
0.51($\pm$ 0.04) & 
1.3($\pm$ 0.2) \\

3.5 & 1.75 & 
1.49($\pm$  0.14) &
0.38($\pm$ 0.04) & 
1.5($\pm$ 0.2) \\

4.0 & 2.0 & 
1.43($\pm$ 0.13) &
0.60($\pm$ 0.06) &
1.3($\pm$ 0.2) \\

4.5 & 2.25 & 
1.64($\pm$ 0.15) & 
0.64($\pm$ 0.03) & 
1.3($\pm$ 0.3) \\

5.0 & 2.5 & 
1.34($\pm$  0.19) &
0.50($\pm$ 0.04) &
1.4($\pm$  0.2)  \\

\hline
\end{tabular}
\caption{
The optimized indices for the bandwidth distribution 
and level spacing distribution on the critical lines 
of $\mu = 1$ and $\lambda=2\mu$.
For the definition of $\gamma$,
see Eq.(\ref{tail2}). 
}
\label{table;not_l2}
\end{table}

{\it Acknowledgments} 
Y.T. and K.I. thank S. Hikami for useful discussions. 
K.I. thanks A. Garcia-Garcia for discussion
and 
M. Kohmoto for explaining previous results. 
K.I. has benefited from 
the Grand-in-Aid for Science(B), No.1430114  of JSPS.

\end{document}